\pdfoutput=1
\documentclass[11pt]{article}
\usepackage[dvipsnames]{xcolor}
\usepackage[final]{acl}
\usepackage{listings}
\usepackage{multirow}
\usepackage{tabularray}
\usepackage{hyperref}
\usepackage{times}
\usepackage{amsmath}
\usepackage{latexsym}
\usepackage{booktabs}
\usepackage{float}
\usepackage{tcolorbox}
\usepackage[T1]{fontenc}
\usepackage[utf8]{inputenc}

\usepackage{microtype}
\tcbuselibrary{listings, skins, breakable, theorems}
\usepackage{inconsolata}

\usepackage{graphicx}
\title{
How effective are  VLMs in assisting humans in inferring the quality of mental models from Multimodal short answers?}
\author{
  Pritam Sil\textsuperscript{1} 
  Durgaprasad Karnam\textsuperscript{2} 
  Vinay Reddy Venumuddala\textsuperscript{3} 
  Pushpak Bhattacharyya\textsuperscript{1} \\
  \textsuperscript{1}Department of Computer Science and Engineering, IIT Bombay, Mumbai, India \\
  \textsuperscript{2}Center for Educational Technology, IIT Bombay, Mumbai, India \\
  \textsuperscript{3}School of Management, Mahindra University, Hyderabad, India \\
  \texttt{pritamsil@cse.iitb.ac.in, karnamdpdurga@gmail.com} \\
  \texttt{vinay.venumuddala@mahindrauniversity.edu.in} \\
}

\begin{document}
\maketitle
\begin{abstract}
STEM Mental models can play a critical role in assessing students’ conceptual understanding of a topic. They not only offer insights into what students know but also into how effectively they can apply, relate to, and integrate concepts across various contexts. Thus, students' responses are critical markers of the quality of their understanding and not entities that should be merely graded. However, inferring these mental models from student answers is challenging as it requires deep reasoning skills. We propose \textit{MMGrader}, an approach that infers the quality of students' mental models from their multimodal responses using concept graphs as an analytical framework. In our evaluation with 9 openly available models, we found that the best-performing models fall short of human-level performance. This is because they only achieved an accuracy of approximately 40\%, a prediction error of 1.1 units, and a scoring distribution fairly aligned with human scoring patterns. With improved accuracy, these can be highly effective assistants to teachers in inferring the mental models of their entire classrooms, enabling them to do so efficiently and help improve their pedagogies more effectively by designing targeted help sessions and lectures that strengthen areas where students collectively demonstrate lower proficiency.\end{abstract}
  
\section{Introduction}
\label{Introduction}
Learning Science, Technology, Engineering, and Mathematics (STEM) requires students to build and manipulate abstract entities, such as mental models\footnote{We use the terms ``mental models" and ``STEM mental models" synonymously throughout this work.}. Mental models are ``basic units of coherently structured knowledge'' ~\citep{hestenes2009modeling}, representing cognitive structures that represent scientific understanding. They are often embodied in what is called physical intuition. However, students do not always construct coherent mental models, which can limit their conceptual understanding. Consequently, assessing the quality of students’ mental models can serve as a proxy for evaluating the depth of their learning in STEM topics. Inferring such mental models at scale, particularly from handwritten student responses, presents a significant challenge.  

Historically, educational measurement has relied on rubric-based approaches for evaluating student responses \cite{Hillegas1912}.  However, such approaches are often labour-intensive, requiring instructors to create separate rubrics for each question, which limits scalability and generalizability.  Recent advances in Artificial Intelligence (AI) offer new possibilities for inferring mental models from open-ended student responses. Natural Language Processing (NLP) techniques enable semantic similarity analysis, automated concept extraction, and discourse-level modelling, allowing evaluation beyond surface-level scoring. While these approaches can generate similarity scores, they do not explicitly reveal the underlying mental model of a student.  

To address these limitations, we propose shifting the focus from scoring individual answers to directly inferring students’ mental models as a measure of learning quality. In this exploratory study, we investigate whether students' underlying STEM mental models can be inferred from multimodal handwritten responses, which often contain both text and diagrams. This task requires deep reasoning capabilities, whether performed by humans or AI. Inferring mental models in formal STEM domains is challenging, as it involves higher-order cognitive processes that typically require specialised expertise even among human annotators.  

\newpage
Our contributions are as follows -
\begin{enumerate}
    \item \textit{MMGrader:} An approach for inferring students’ STEM mental models from multimodal responses. (Section~\ref{STEM3Grader: An Algorithm to Infer mental models})  
    \item MMGrader decomposes topics into fundamental units, called concept links, and evaluates student responses using a provided mapping between questions and concept links, along with a predefined scoring scale. (Section~\ref{STEM3Grader: An Algorithm to Infer mental models})   
    \item Extensive Evaluation of how close are openly available vision–language models (VLMs) to approximating human annotators in reasoning over students’ multimodal responses to infer mental models. (Section~\ref{Experimental Evaluation of existing VLMs})
    \item Evaluations reveal that the best performing model achieves an accuracy of \~40\% along a prediction error of 1.1 units, and a scoring distribution fairly aligned with human scoring patterns. (Section~\ref{Results})

\end{enumerate}

\section{Related Work}
\label{Related Work}
 Kenneth Craik is credited with introducing the term ``mental models''. In education, mental models serve as an effective tool for understanding a student's proficiency in a topic. However, researchers have quantified these models in various ways, leading to diverse approaches for assessing student understanding.  

Based on this concept, \citet{RusLinteanAzevedo2009} inferred whether a student possesses a high, intermediate, or low mental model from student-generated paragraphs written during prior knowledge activation, a self-regulatory process. The answers came from the topic of the circulatory system in biology. Their approach was evaluated using traditional machine learning techniques such as Naive Bayes (NB), Bayes Nets (BNets), Support Vector Machines (SVM), Logistic Regression (LR), and two variants of decision trees. While effective, this line of work focused on only labels, and thus, a more detailed structure of mental models is still necessary for complex tasks such as grading.  

To capture more detailed representations, \citet{maharjan2019concept} used concept maps as a tool for modelling students' mental models. They proposed a novel method called DT-OpenIE to extract tuples from student answers. This was achieved by converting sentences into short clauses and then extracting tuples, which were compared to ground-truth tuples provided by human annotators to grade student answers. Although this approach provided more structured insights, its dataset contained only short textual answers (1–2 sentences), which does not reflect the broader variety of student responses. 

Extending this direction, \citet{agarwal-etal-2022-multi} introduced the use of Abstract Meaning Representation (AMR) graphs. In these graphs, nodes represent concepts or predicates, while edges capture relations such as subject/object. These elements were extracted from the student's answer to construct the AMR graph. The same is performed on the reference answer, converted to embeddings and compared to generate the final score. This approach, again, can be seen as inferring mental models from student answers. Despite this detailed methodology, the answers in this study were also limited to textual responses. 

Similarly, \citet{sahu2025directed} proposed constructing answer graphs that resemble AMR graphs, thereby representing students’ mental models in a structured way to identify gaps in student responses. Once again, however, the answers considered were short (1–2 sentences) and textual.  

Apart from short textual answers, \citet{10.1145/3597926.3598049} extracted concept graphs from code submitted as answers to CS-1 programming assignments. These graphs were compared with those derived from reference solutions to automatically grade programming answers. While this demonstrates the applicability of concept graph–based approaches to other domains, it still does not address multimodal or longer responses.  

Prior work has employed a range of representations for mental models, including simple labels (high/intermediate/low), concept maps, concept graphs, and AMR graphs. However, these approaches share several limitations as they primarily focus on short textual answers. Moreover, they do not account for symbols and notations that are common in domains such as physics, and they lack generalizability across disciplines. To address these challenges, we propose \textit{MMGrader}, a novel approach for inferring mental models from multimodal student answers using concept graphs as an analytical framework. We begin by explaining how mental models can be represented in educational contexts in the next section.

\section{How to represent mental models in Education ?}
\label{How to represent mental models in Education ?}
As mentioned by \citet{hestenes2009modeling}, mental models are ``units of coherently structured knowledge'' formed in the human mind when interacting with real-world objects. They can be ``directly compared with physical things and processes'' and are ``embodied in physical intuition.'' This raises a fundamental question: how can we represent them? In the context of education, one effective representation is through concept graphs.  

\subsection{Concept Hierarchy}
\begin{figure}[H]
    \centering
    \includegraphics[width=\linewidth]{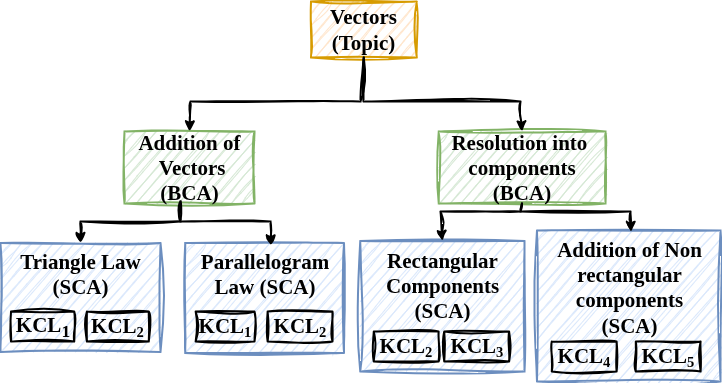}
    \caption{Concept hierarchy for the topic of vectors}
    \label{Topic Tree}
\end{figure}

In STEM education, each topic can be organised according to a concept hierarchy. A topic can be divided into broader concept areas (BCAs) or subtopics, which are further refined into sub-concept areas (SCAs). SCAs, in turn, contain fundamental units called Key Concept Links (KCLs). When an educator teaches a topic or subtopic, these KCLs serve as the learning outcomes for that topic. A coherent mental model acquired by a student revolves around these KCLs.

Figure~\ref{Topic Tree} illustrates a concept hierarchy for the topic of vectors. For example, when an educator explains the triangle law of vector addition, the learner strengthens their understanding of the direction ($\text{KCL}_1$) and magnitude ($\text{KCL}_2$) of vectors. Once incorporated into their mental model, these KCLs enable students to reason about how an object will move when multiple forces act upon it.  

\subsection{Concept Graph}
\begin{figure}[H]
    \centering
    \includegraphics[width=0.9\linewidth]{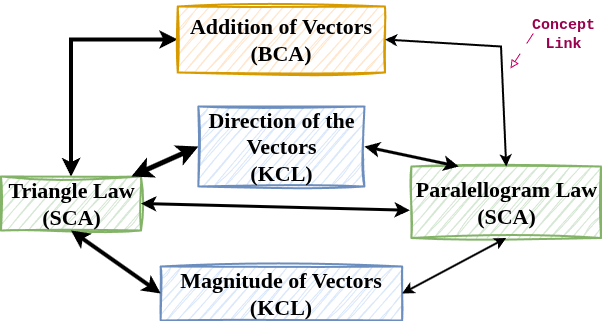}
    \caption{A concept graph for addition of vectors}
    \label{Concept_graph}
\end{figure}

Building on the idea of concept hierarchies, we adopt the methodology of \citet{karnam2021touchy} to define a concept graph $G(V,E)$, where the vertices $V$ are conceptual units and the edges $E$ are concept links. The set of vertices includes components from the concept hierarchy: $V = BCA \cup SCA \cup KCL$. Each edge represents a relationship between two nodes. For instance, the link between the KCL “direction of vectors” and the “triangle law” reflects that understanding the triangle law strengthens a student’s grasp of vector direction.  

\subsection{Concept Graphs as tools to represent mental models}
Mental models are not directly accessible but are indicated by markers. Identifying such markers present in student responses is a challenging task. We present concept graphs as analytical frameworks to effectively perform this task.  

This is because concept graphs capture strong associations between concepts within a topic, aligning closely with the definition of mental models as ``units of coherently structured knowledge.''  Each edge in the graph can be assigned a strength score, reflecting how well a student has learned a particular concept. A higher score indicates more coherent and high-quality mental models and a greater understanding, which in turn suggests that the student can successfully apply that concept across other contexts.  

Moreover, the concept graphs used for a particular topic need not be unique. An instructor can easily customise them according to the learning outcomes of a course. They can be seen as data structures to infer the completeness, correctness and strength of the underlying mental models. Building on this idea, we propose a new approach called MMGrader to infer students' mental models from multimodal responses.

\section{MMGrader: An approach to infer students' mental models}
\label{STEM3Grader: An Algorithm to Infer mental models}

Concept graphs can be effective analytical frameworks to infer students' mental models from their responses, as demonstrated by \citet{karnam2021touchy}. Using these analytical frameworks on students' multimodal responses requires deep expertise and reasoning capacities, even among humans.  By multimodal, we mean that the answers may involve text, diagrams, or a combination of the two. MMGrader addresses this by employing an analytical framework that involves breaking down the topic into concept links, establishing a relevancy mapping between questions and the concept links, and providing a rubric for evaluation. The approach is designed in such a way that instructors do not need to prepare any additional input beyond what is already present in standard STEM education.

\begin{figure*}[h!tbp]
    \centering
    \includegraphics[width=\linewidth]{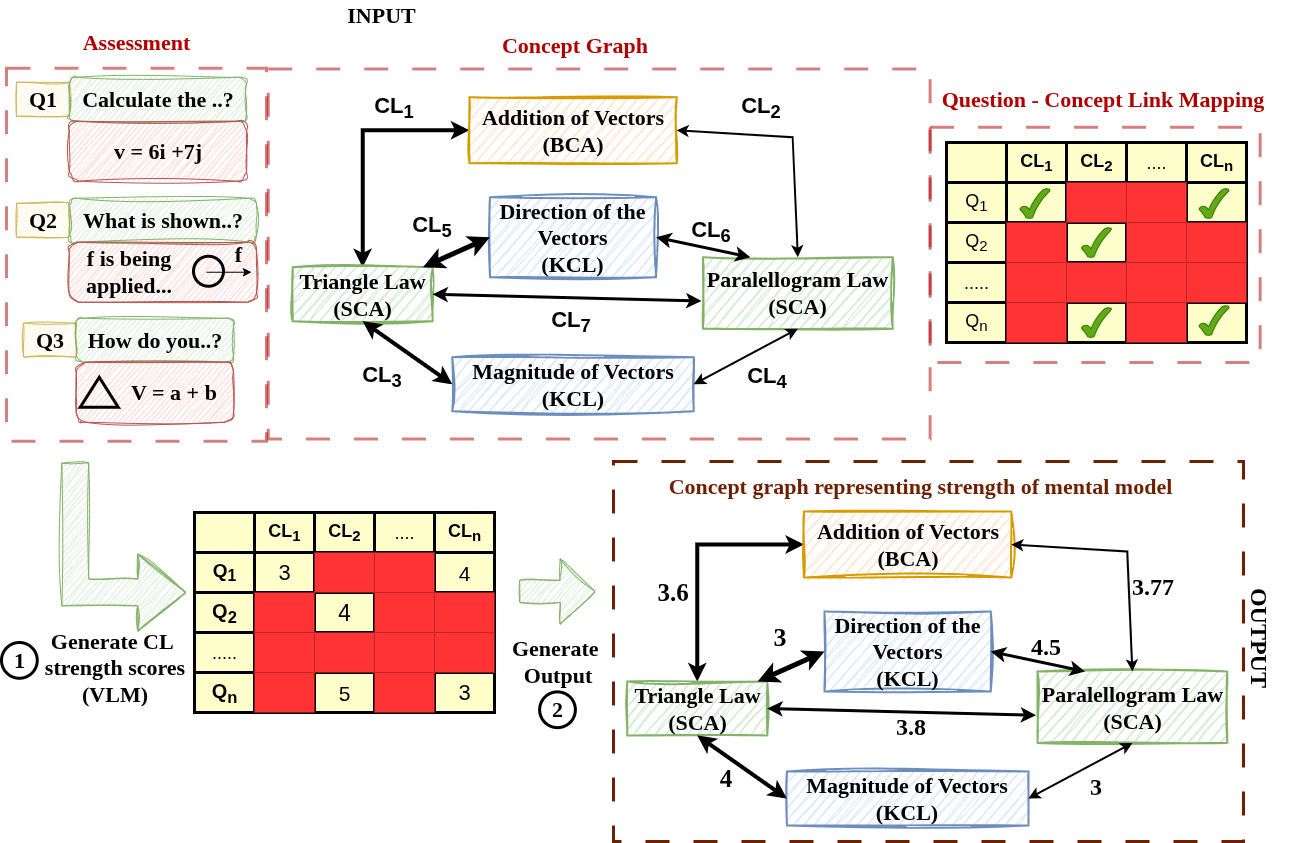}
    \caption{Overview of MMGrader}
    \label{STEM3Grader}
\end{figure*}
The details of  MMGrader\footnote{Our code is available at : \href{https://github.com/psil123/MMGrader}{https://github.com/psil123/MMGrader}}  (Figure~\ref{STEM3Grader}) are as follows:-

\textbf{Input:} MMGrader takes in the following as the input:-  
\begin{itemize}
    \item \textit{Student assessment:} A student assessment containing handwritten responses containing text and diagrams.  
    \item \textit{Concept graph:} The concept graph associated with the topic. It serves as an analytical framework to analyse student responses.
    \item \textit{Question–concept link mapping:} A mapping between a question to its relevant concept links.
\end{itemize}
Each topic includes a predefined concept hierarchy that is typically available and standardised in STEM curricula. During assessment design, the instructor maps each question to its relevant concept links. These concept links can be directly linked to the learning outcomes of a topic. Hence, the approach is flexible and can customized according to the instructor's requirements.   

\textbf{Step 1 (Evaluation):} Given the question–concept link mapping, the proposed approach evaluates student answers to estimate a strength score for each concept link associated with a question. This is a challenging task as it requires diagram understanding capabilities along with deep reasoning skills. While humans can perform this task easily, we investigate whether openly available VLMs can replicate this ability.  

\textbf{Step 2 (Generation):} Using the VLM-generated strength scores, the proposed approach constructs a concept graph representing the strength of the student’s mental model. To generate the strength scores, we adopt a bottom-up approach along the concept hierarchy, starting from KCL–SCA pairs. The resultant strength score is computed as the average of all scores shown by the student across relevant questions. For example, say $\text{CL}_{2}$ (Figure~\ref{STEM3Grader}) is being measured in question 1 and question 2, then its strength is computed as the average of the scores obtained in both. Similarly, we compute the strength scores of other CLs as average of the scores obtained on their mapped questions. The intuition is that once a student has grasped a fundamental concept, that is, a concept link, they should apply it consistently across all their answers. Similarly, we compute the other strength scores. This concept graph, built with the strengths of each link for a student, stands as a representation of the student's mental model. 

The performance of MMGrader depends on the diagram interpretation and reasoning capabilities of the underlying VLMs. Therefore, evaluating MMGrader is equivalent to evaluating the performance of VLMs, details of which are provided in the next section.

\section{How close are VLMs to Humans ?}
\label{Experimental Evaluation of existing VLMs}
MMGrader leverages concept graphs as analytical frameworks to represent a student's STEM mental models. As a result, the task reduces to a simple scoring task. The scoring involves assigning a strength score to a concept link given the question, answer and scoring guidelines. To evaluate whether VLMs are comparable to human annotators on this task, we first construct a dataset as described in the next section.

\subsection{Dataset Construction}
\begin{figure*}[h]
    \centering
    \includegraphics[width=\linewidth]{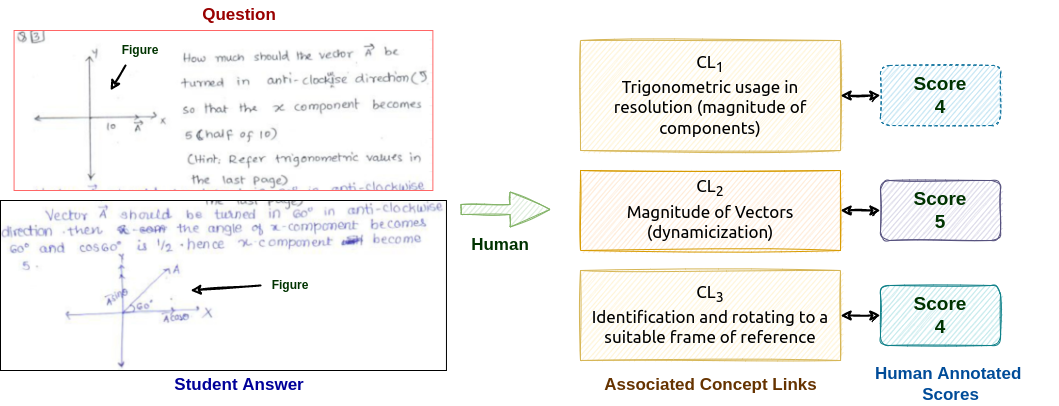}
    \caption{Sample from our dataset}
    \label{sample}
\end{figure*}
The dataset was collected from an assessment consisting of \textit{10 questions}, jointly prepared by educators and education researchers. The assessment was attempted by \textit{6 students}, each of whom had varying scholastic levels, and the answer sheets were collected and annotated by 6 human annotators. All human annotators have at least 2-4 years of experience in education practice or research, and were working in a premier center of education research. 

Each question was manually mapped to a subset of \textit{12 concept links} on the topic of addition of vectors and resolution of vectors from the 11th-grade curriculum. 
To establish ground-truth scores, each annotator was responsible for rating 2 answer sheets. After completing their individual annotations, all annotators met to resolve discrepancies in the assigned scores through offline discussion over the provided rubrics to reach a consensus. As a result, each \textit{question–concept link–student answer} triplet was assigned a ground-truth strength score. The final dataset contained \textit{895 data points}, where each data point corresponds to such a triplet with its associated score. This controlled annotation process provides more reliable ground-truth labels and ensures that the gathered data is of high quality.

All questions and answers in the dataset were handwritten. Each question included a diagram, while answers could be text-only, diagrammatic, or a combination of both. To extract the handwritten text from questions and answers, we used Google’s Gemini model (\texttt{gemini-2.5-flash}) for optical character recognition (OCR). Gemini has been shown to outperform existing models on several OCR benchmarks\footnote{\url{https://benchmarking.nanonets.com/} (Last accessed: 06/10/2025)}. Figure~\ref{sample} shows a single data point from the dataset.  

We plan to expand this dataset with additional samples collected from real-world examinations across various subjects, thereby increasing its scale and complexity and enabling a broader evaluation of multimodal student assessments.  

\subsection{Experimental Evaluation}
\label{Experimental Evaluation}
\label{Results}
% \begin{table*}
% \centering
% \begin{tabular}{p{1cm}cccccccccccc}
% \toprule
% \multirow{2}{*}{Model} & \multicolumn{3}{c}{Base Scoring Scenario} & \multicolumn{3}{c}{Generic Scoring Scenario} & \multicolumn{3}{c}{Detailed Scoring Scenario} & \\
% \cmidrule(r){2-4} \cmidrule(r){5-7} \cmidrule(r){8-10} \cmidrule(r){11-13}
%  & \textbf{Accuracy} & \textbf{RMSE} & \textbf{EMD} & \textbf{Accuracy} & \textbf{RMSE} & \textbf{EMD}& \textbf{Accuracy} & \textbf{RMSE} & \textbf{EMD}\\
% \midrule
% Molmo    &      29.66 & 1.42 & 0.08 &    \textbf{39.98} &  \textbf{1.1} & \textbf{0.12}  &    \textbf{32.5} & \textbf{1.26} & \textbf{0.03} \\
% Pixtral      &  \textbf{29.82} & \textbf{1.5} & \textbf{0.04} &  29.82 & 1.5 & 0.04 &  28.7 & 1.45 & 0.05 \\
% InternLM  & 20.4 & 1.78 & 0.15& 26.93 & 1.66 & 0.12 & 22.78 & 1.68 & 0.15 \\
% Gemma     &     22.54 & 1.66 & 0.16 &    24.44 &  1.6 & 0.12 &  18.62 &  1.7 & 0.19\\
% LLama-Vision   & 19.77 & 1.98 & 0.19 &  24.27 & 1.7 & 0.1  &  22.71 & 1.71 & 0.11\\
% Gemini  &  23.01 & 1.6 & 0.14 &  22.89 & 1.6 & 0.13 &  19.95 & 1.66 & 0.15\\
% Qwen    & 18.15 & 1.83 & 0.25  & 18.15 & 1.83 & 0.25&  26.33 & 1.51 & 0.13\\
% LLava      &    13.17 &  2.0 & 0.25 &    13.76 & 1.95 & 0.24 &  13.4  &1.99 & 0.24\\
% Granite     &     8.9 & 2.25 & 0.24 &    10.79 & 2.24 & 0.26 &  9.37  & 2.2 & 0.23\\

% \bottomrule
% \end{tabular}
% \caption{Performance of openly available VLMs on our dataset}
% \end{table*}

\begin{table*}
\centering
\begin{tabular}{p{1cm}cccccccccccc}
\toprule
\multirow{2}{*}{Model} & \multicolumn{3}{c}{Base Scoring Scenario} & \multicolumn{3}{c}{Generic Scoring Scenario} & \multicolumn{3}{c}{Detailed Scoring Scenario} & \\
\cmidrule(r){2-4} \cmidrule(r){5-7} \cmidrule(r){8-10} \cmidrule(r){11-13}
 & \textbf{Accuracy} & \textbf{RMSE} & \textbf{EMD} & \textbf{Accuracy} & \textbf{RMSE} & \textbf{EMD}& \textbf{Accuracy} & \textbf{RMSE} & \textbf{EMD}\\
\midrule
Molmo    &      29.66 & 1.42 & 0.42 &    \textbf{39.98} &  \textbf{1.1} & \textbf{0.58}  &    \textbf{32.5} & \textbf{1.26} & \textbf{0.17} \\
Pixtral      &  \textbf{29.82} & \textbf{1.5} & \textbf{0.45} &  29.82 & 1.5 & 0.22 &  28.7 & 1.45 & 0.49\\
InternLM  & 20.4 & 1.78 & 0.74& 26.93 & 1.66 & 0.58 & 22.78 & 1.68 & 0.77 \\
Gemma     &     22.54 & 1.66 & 0.8 &    24.44 &  1.6 & 0.61 &  18.62 &  1.7 & 0.93\\
LLama-Vision   & 19.77 & 1.98 & 0.93 &  24.27 & 1.7 & 0.5  &  22.71 & 1.71 & 0.53\\
Gemini  &  23.01 & 1.6 & 0.72 &  22.89 & 1.6 & 0.69 &  19.95 & 1.66 & 0.75\\
Qwen    & 18.15 & 1.83 & 3.89  & 18.15 & 1.83 & 0.91&  26.33 & 1.51 & 0.64\\
LLava      &    13.17 &  2.0 & 1.23 &    13.76 & 1.95 & 1.21 &  13.4  &1.99 & 1.21\\
Granite     &     8.9 & 2.25 & 1.18 &    10.79 & 2.24 & 1.3 &  9.37  & 2.2 & 1.13\\

\bottomrule
\end{tabular}
\caption{Performance of openly available VLMs on our dataset}
\end{table*}

The effectiveness of  MMGrader depends on how well VLMs can assign strength scores on various concept links. To determine this, we evaluate a total of 9 openly available models that support multiple images along with complex prompts. Out of these models, Granite~\citep{GraniteVision2025} is a small 2B model while Molmo~\citep{MolmoAndPixMo2024}, Qwen~\citep{Qwen2VL2024} and LLaVa~\cite{Qwen2VL2024} are 7B models. Apart from this, we have LLamaVision~\citep{liu2023llava}, which is an 11B model and Pixtral~\citep{agrawal2024pixtral} and Gemma~\citep{Gemma32025}, which are 12B models. We have also considered Gemini\footnote{\url{https://docs.cloud.google.com/gemini-enterprise-agent-platform/models/gemini/2-5-flash} (Last Accessed: 12/08/2026)}, as it is closed-source but openly available. More details have been added in Appendix~\ref{VLMs in Consideration}. The main objective of this set of experiments is to determine whether existing VLMs are capable of evaluating handwritten multimodal answers and inferring the underlying mental models. 

Three different experimental settings were considered, for which the detailed prompts can be found in Appendix~\ref{Prompts Used}. The experimental settings were as follows:  
\begin{itemize}
    \item \textbf{Base Scoring Scenario:} The VLM is instructed to generate an integer strength score between 1 and 5, without specifying what each score represents. The prompt required the models to rely entirely on prior knowledge from pretraining when evaluating answers. 

    \item \textbf{Generic Scoring Scenario:} The VLM is instructed to generate an integer strength score between 1 and 5, with generic specifications provided for each score. This setting offers a more guided scale, but the model must still interpret the concept link, question, and answer based on its prior knowledge. The scoring rubric added to the prompt  was:  
    \begin{itemize}
        \item 1: No indication of ability to handle the link  
        \item 2: Very limited familiarity with the concept  
        \item 3: Inconsistent procedure, reproducing textbook understanding without modification  
        \item 4: Partial conceptual understanding, applying with some deviations from standard textbook usage  
        \item 5: Strong conceptual understanding  
    \end{itemize}  

    \item \textbf{Detailed Scoring Scenario:} The VLM is instructed to generate an integer strength score between 1 and 5 with a concept-link–specific rubric. For example, for the concept link \textit{magnitude of vectors}, the scoring scale used was:  
    \begin{itemize}
        \item 1: No indication of understanding regarding the manipulation of magnitudes  
        \item 2: Very limited understanding of magnitudes  
        \item 3: No ability to dynamically simulate vectors  
        \item 4: Partial conceptual understanding of vector manipulation  
        \item 5: Strong conceptual understanding of vector manipulation  
    \end{itemize}  
    This was the most guided setting, where each score definition was explicitly tied to the concept link, and the model’s task was to map the student’s answer to the appropriate score. This is the exact scenario which were provided to humans during the scoring task. 

    \item \textbf{Chain-of-Thought (CoT) Scoring Scenario:} CoT is a prompting technique introduced by \cite{wei2022chainofthought}, which is effective in encouraging models to perform detailed reasoning before responding to a task. In our CoT prompt, we provide a generic scoring scale along with detailed reasoning steps. This set of experiments was designed to evaluate whether model performance improves when the models are explicitly instructed on how to think before generating the strength score.

\end{itemize}  

We evaluated model performance using three metrics: exact match accuracy, root mean squared error (RMSE), and earth mover’s distance (EMD)~\citet{rubner1998metric}. Exact match accuracy measures whether the model’s score exactly matches the human-annotated score, with higher values indicating closer alignment with human judgment. RMSE quantifies the deviation of model scores from human scores, while EMD evaluates whether the distribution of scores produced by the model resembles that of humans. A lower RMSE or EMD indicates closer alignment in scoring patterns.  

Across all experiments, the best performance was achieved by \textbf{Molmo}, with an accuracy of 39.98\%, RMSE of 1.1, and EMD of 0.58. This suggests that Molmo’s predictions were typically within one unit of the human-assigned score and that its overall scoring distribution was fairly close to that of human annotators. We attribute this to Molmo’s training on educational data, which has also enabled it to outperform other models on educational benchmarks.  

\textbf{Pixtral} consistently achieved a low EMD score across all experimental scenarios, indicating that its scoring distribution closely resembled human scoring. However, with an RMSE of approximately 1.5, its predictions tended to deviate further from the ground truth on an individual level, resulting in lower accuracy. This suggests that Pixtral could easily mimic human scoring patterns if fine-tuned on this data. In comparison, \textbf{InternLM} also performed reasonably well, likely due to its pretraining on scientific notations and tokens. However, its performance lagged behind Molmo because of overthinking and other reasoning-related issues, as discussed in Section~\ref{Ablation Studies}. \textbf{Gemma} and \textbf{LLamaVision} followed a similar trend.  

For \textbf{Qwen}, we observed consistent improvements in performance as more fine-grained details were added to the prompt. Although Qwen is not specifically trained on educational data, it is able to pick up the additional details provided in the prompt and use them to evaluate the answers.  

\textbf{Gemini} showed relatively strong performance, particularly in interpreting handwritten diagrams. However, its reasoning abilities sometimes introduced unnecessary confusion and overthinking into scoring decisions. This was evident in its generated rationales, as discussed in Section~\ref{Ablation Studies}.  

\textbf{LLaVa}, trained primarily on general-purpose data rather than educational datasets or hand-drawn diagrams, was limited in its performance on this task.  

Finally, \textbf{Granite} performed the worst overall. As a smaller model trained on relatively simple tasks and without any prior hand-drawn images in its training data, its predictions were the least aligned with human annotations. This also explains why it frequently repeats the same sentence in its outputs.

\begin{table}[h!tbp]
\begin{tabular}{cccc}
\toprule
     &   \textbf{Accuracy} & \textbf{RMSE} & \textbf{EMD} \\
     \midrule
Molmo      &    \textbf{34.16} & \textbf{1.27} & \textbf{0.63}\\
Qwen     &      29.66 & 1.44 &  0.6\\
InternLM  &   25.27 & 1.62 & 0.53\\
Pixtral    &    24.32 & 1.65  &0.48\\
Gemini    &     23.13 & 1.61 & 0.58\\
Gemma       &    20.4 & 1.66 & 0.81\\
LLamaVision &   20.47 & 1.97 & 0.78\\
LLava     &     10.56 & 2.09 & 1.19\\
Granite   &      0.95 & 3.93 & 3.63\\

\bottomrule
\end{tabular}
\caption{Performances of VLMs on CoT-based scenario}
\label{cot}
\end{table}

% \begin{table}[]
% \begin{tabular}{cccc}
% \toprule
%      &   \textbf{Accuracy} & \textbf{RMSE} & \textbf{EMD} \\
%      \midrule
% Molmo   & \textbf{39.74} & \textbf{1.1} & \textbf{0.12} \\
% Qwen  &  29.66 & 1.44 & 0.12\\
% Pixtral  &  28.7 & 1.45 & 0.05\\
% InternLM  &  25.27 & 1.62 & 0.11\\
% Gemini  &  23.13 & 1.61 & 0.12\\
% LLamaVision & 20.47 & 1.97 & 0.16 \\
% Gemma        &  20.4 & 1.66 & 0.16\\
% LLava        & 10.56 & 2.09 & 0.24\\
% Granite   &  0.95 & 3.93 & 0.73\\
% \bottomrule
% \end{tabular}
% \caption{Performs of models on CoT based prompting}
% \label{cot}
% \end{table}

We also performed a similar set of experiments using a chain-of-thought (CoT) prompting strategy (Table~\ref{cot}). The results followed the same trend, further validating our earlier observation that \textbf{Molmo} is the most effective candidate among all VLMs. One interesting observation is that \textbf{Qwen} exhibited improved performance when provided with more fine-grained details. This suggests that, Qwen is able to leverage the step-by-step reasoning cues to perform better on the task.

To summarise, openly available VLMs still fall short of human-level performance at this task. With improved accuracy, these can be highly effective assistants to teachers in inferring the mental models of their entire classrooms, enabling them to do so efficiently and help improve their pedagogies more effectively by designing targeted help sessions and lectures that strengthen areas where students collectively demonstrate lower proficiency.

\subsection{Qualitative Analysis of VLM Outputs}
\label{Ablation Studies}
While the quantitative results provide an overall picture of the current state of existing VLMs, we also investigate why certain VLMs perform poorly.

When examining the predicted values for InternLM, Qwen, and Pixtral, we observed that they produced incorrect strength scores. We conjecture that this behaviour arises from overthinking and self-contradiction. An illustrative example from InternLM is shown in the box below.

\begin{tcolorbox}[breakable, colback=blue!5!white, colframe=gray!75!black, title=Output generated by InternLM, label=internlm_output_box]
...

First, I need to understand what the vectors $\vec{P_1}$ and $\vec{P_2}$ represent. 
The notation here seems a bit confusing. Typically, when vectors are written like 
$\vec{P_1} = 6 \cos(50^\circ)$ and $\vec{P_2} = 6 \sin(50^\circ)$, 
they might be components of a vector in a coordinate system.

...

Wait, maybe $\vec{P_1}$ and $\vec{P_2}$ are unit vectors in the $x$ and $y$ directions multiplied by 6. 

...

But the original statement just says $\vec{P_1} + \vec{P_2} = \vec{R}$. If $\vec{P_1}$ and $\vec{P_2}$ are components of $\vec{R}$, then yes, their sum would equal $\vec{R}$. However, the notation here is unclear. 

...

Wait, perhaps the question is using a different notation where $\vec{P_1}$ and $\vec{P_2}$ are components of a vector in polar coordinates.

...
\end{tcolorbox}

\begin{figure}
    \centering
    \includegraphics[width=0.9\linewidth]{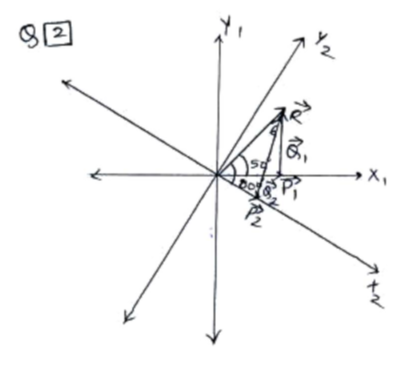}
    \caption{Image provided as part of one of the questions in the dataset.}
    \label{ablation_q_internlm}
\end{figure}

The predicted sentences indicate that InternLM is not incorporating the given diagram in the question into its reasoning process, even though it provides a clear hint regarding what $\vec{P_1}$ and $\vec{P_2}$ actually represent (Figure~\ref{ablation_q_internlm}).

For LLaVa, the model only predicted either 1 or 5 as the strength score, which corresponds to examples of the output format provided in the prompt. This indicates that LLaVa is not interpreting the prompt correctly. Similarly, smaller VLMs such as Granite also failed to interpret the prompt, suggesting that it is mainly trained for simpler tasks such as caption generation. As a result, it generated random strength scores. In one instance where it failed to generate any value, it instead produced the following: \texttt{<start of description> The image consists of a geometric diagram with a central point labelled as "O". From this point, three vectors are drawn, labelled as "B", "C", and "Y"}. This again demonstrates its inability to interpret hand drawn diagrams (Figure~\ref{ablation_q_internlm}), highlighting the need to finetune it on handwritten data paired with simple instructions.

In the case of closed models, Gemini was able to generate strength scores with an accuracy of ~23\% for all scenarios. However, instead of following the prescribed output format, it often returned numbers in the form \verb|\boxed{3}|, which is consistent with conventions used in standard benchmarks. While valid in that context, this behaviour is undesirable when the model is explicitly instructed to follow a different format.

\section{Conclusion}
Students' conceptual understanding in STEM depends on the coherence, accuracy and completeness of underlying STEM mental models. Inferring their mental models is hence a more effective way to design effective pedagogies. Students' responses are not just things to be graded, but as critical markers of the quality of their understanding. We present a method (MMGrader) that addresses the problem of inferring mental models using concept graphs as frameworks to analyse the multimodal responses.

Our evaluations indicated \textit{Molmo} as the best-performing model across all parameters, even though there is a 1.1-unit difference between Molmo and expert scores. These results suggest a promising direction for utilising VLM-based approaches in designing scalable and interpretable auto-grading systems that can effectively quantify student understanding. This opens avenues for areas such as personalised education and Intelligent Teaching Assistants.

\section*{Limitations}
The VLMs still require training on a real-life dataset to enhance their accuracy and bring them up to par with human annotators. 
\section*{Ethical Considerations}
The dataset was collected with prior permission from the students. The identities of both the students and annotators had been kept anonymous during the whole process.

\bibliography{custom}
\appendix
\begin{table*}[h]
    \centering
    \begin{tabular}{ccc}
    \toprule
    \textbf{Model Name} & \textbf{Variant} & \textbf{Size}  \\ \midrule
Granite & \texttt{ibm-granite/granite-vision-3.3-2b} & 2B \\
Molmo & \texttt{allenai/Molmo-7B-D-0924} & 7B \\
Qwen & \texttt{Qwen/Qwen2-VL-7B-Instruct} & 7B\\ 
LLaVa & \texttt{llava-hf/llava-v1.6-mistral-7b-hf} & 7B \\
Interlm & \texttt{internlm/Intern-S1-mini} & 8.54B\\
LLamaVision & \texttt{meta-llama/Llama-3.2-11B-Vision-Instruct} & 11B\\ 
Pixtral & \texttt{mistralai/Pixtral-12B-2409} & 12B \\
Gemma & \texttt{google/gemma-3-12b-it} & 12B \\
Gemini & \texttt{gemini-2.5-flash} & -\\
\bottomrule
    \end{tabular}
    \caption{Configuration of VLMs used}
    \label{VLMSused}
\end{table*}
\section{Prompts Used}
\label{Prompts Used}
Below are the prompts used for various scenarios. Note that for the Detailed Scoring Scenario, we replace the scoring scale with the concept link-specific details as shown in Section~\ref{Experimental Evaluation}. 

\begin{tcolorbox}[breakable, colback=blue!5!white, colframe=gray!75!black, title=Base Scoring Prompt, label=base]
You are an expert evaluator of student responses. 

The provided questions and student answers belong to the topic of addition and resolution of vectors from 11th standard. 

The first image provided belongs to the question.

A second image if present belongs to the student answer.

A strength score is a number between 1 and 5 (both inclusive) which is used to represent how well a concept link has been expressed in the student answer.

Task : Your task is to generate the strength score of the concept link by analyzing the question and student answer pair.

Output Format (strict):

<Score>an integer between 1 and 5</Score>

Examples:

<Score>1</Score>

<Score>5</Score>
\end{tcolorbox}

\begin{tcolorbox}[breakable, colback=blue!5!white, colframe=gray!75!black, title=Generic Scoring Scenario, label=generic]
You are an expert evaluator of student responses. 

The provided questions and student answers belong to the topic of addition and resolution of vectors from 11th standard. 

The first image provided belongs to the question.

A second image if present belongs to the student answer.

A strength score is a number between 1 and 5 (both inclusive) which is used to represent how well a concept link has been expressed in the student answer.

Task : Your task is to generate the strength score of the concept link by analyzing the question and student answer pair.

Strength Score Scale (1–5):

1 : (No indication of ability to handle the link) 

2 : (Very little familiarity with the skill)  

3 : (Inconsistent Procedure) Trying to impose the text book understanding without any modification. 

4 : (Inconsistent Concept/ Procedure- applying with some changes from a regular textbook usage) 

5 : (Strong Conceptual)

Output Format (strict):

<Score>an integer between 1 and 5</Score>

Examples:

<Score>1</Score>

<Score>5</Score>
\end{tcolorbox}

\begin{tcolorbox}[breakable, colback=blue!5!white, colframe=gray!75!black, title=Chain-of-Though based Generic Scoring Prompt, label=CoT]
You are an expert evaluator of student responses. 

The provided questions and student answers belong to the topic of addition and resolution of vectors from 11th standard. 

The first image provided belongs to the question.

A second image if present belongs to the student answer.

A strength score is a number between 1 and 5 (both inclusive) which is used to represent how well a concept link has been expressed in the student answer.

Task Instructions:

1. Review the question carefully.

2. Evaluate the student’s answer.

3. Consider the scoring guide, which maps scores to descriptions.

4. Choose the score (1–5) that best reflects how effectively the student’s answer demonstrates the concept link.

5. Return only the selected score in the specified output format.

Scoring Guide (1–5):

Each line can be read as "score : general description – concept-link specific description"

1 : (No indication of ability to handle the link) 

2 : (Very little familiarity with the skill)  

3 : (Inconsistent Procedure) Trying to impose the text book understanding without any modification. 

4 : (Inconsistent Concept/ Procedure- applying with some changes from a regular textbook usage) 

5 : (Strong Conceptual)

Output Format (strict):

<Score>an integer between 1 and 5</Score>

Examples:

<Score>1</Score>

<Score>5</Score>
\end{tcolorbox}
\section{VLMs in Consideration}
\label{VLMs in Consideration}
Table~\ref{VLMSused} contains the exact model configurations that were used. All the models except Gemini have been accessed via HuggingFace. For Gemini, we have used their API.

\end{document}